\documentclass[12pt,preprint]{aastex}  
							
\usepackage{geometry}
\usepackage[usenames, dvipsnames]{color}
\usepackage{amsmath,amssymb,graphicx,fancyhdr,natbib, wasysym}

\usepackage[center,tiny]{titlesec}

\begin{document}

\section*{Recent Tectonic Activity on Pluto Driven by Phase Changes in the Ice Shell}
\vspace{0.25in}
\centerline{\textbf{Noah P. Hammond$^{1,*}$, Amy C. Barr$^2$, Edgar M. Parmentier$^{1}$}}

\begin{itemize}
\item[$^{1,*}$]Department of Earth, Environmental, and Planetary Sciences \\Brown University, 324 Brook St., Box 1846 \\Providence, RI 02912 (noah\_hammond@brown.edu)
\item[$^2$]Planetary Science Institute, Tucson, AZ, USA
\end{itemize}

\underline{Key Points}
\begin{enumerate}
\item[1)] The partial freezing of a subsurface ocean may be driving recent extensional tectonic activity on Pluto.
\item[2)] If the ice shell is thicker than 260 km and the ocean freezes, dense ice II forms and causes global volume contraction.
\item[3)] The lack of compressional tectonic features suggests that ice II did not form and Pluto's ocean may have survived to present day.
\end{enumerate}

\vspace{0.5in}
\baselineskip=15 pt
\noindent \textbf{Abstract}:  The \textit{New Horizons} spacecraft has found evidence for geologic activity on the surface of Pluto, including extensional tectonic deformation of its water ice bedrock (see \citet{moore2016geology}). 
One mechanism that could drive extensional tectonic activity is global surface expansion due to the partial freezing of an ocean.
We use updated physical properties for Pluto and simulate its thermal evolution to understand the survival of a possible subsurface ocean.
For thermal conductivities of rock less than 3 W m$^{-1}$ K$^{-1}$, an ocean forms and at least partially freezes, leading to recent extensional stresses in the ice shell. 
In scenarios where the ocean freezes and the ice shell is thicker than $260$ km, ice II forms and causes global volume contractions.
Since there is no evidence for recent compressional tectonic features, we argue that ice II has not formed and that Pluto's ocean has likely survived to present day.

\newpage
\baselineskip=20pt   
\parskip=2pt	
\section{Introduction}
Pluto is the largest object in the Kuiper Belt, with a radius $R=1187$ km \citep{stern2015pluto}, and a bulk density of $\bar{\rho} = 1.86$ g/cm$^3$ \citep{stern2015pluto}, suggesting a mixed ice-rock composition \citep{mckinnon1997composition, mckinnon2008structure}. The \textit{New Horizons} flyby of the Pluto-Charon system recently revealed evidence for diverse geologic activity on the surface of Pluto \citep{stern2015pluto}. \citet{moore2016geology} find that while much of the geologic activity seems to involve volatile ices such as nitrogen, carbon monoxide and methane \citep{McKinnon2016convection}, there is also evidence for geologic activity in Pluto's water ice bedrock. The many troughs and scarps that cut across Pluto's surface are often hundreds of kilometers long and up to $4$ km deep. These features have been interpreted as extensional faults \citep{moore2016geology}. The age of these tectonic features is unknown, though many of the graben do not appear to be heavily degraded, indicating they may be geologically young. No obvious signs of strike-slip or compressional tectonic features have been found so far \citep{moore2016geology}.

What could be driving tectonic activity on Pluto? On many icy moons, tidal forces raised on the satellite by their parent planet can drive recent geologic activity (\textit{e.g.} \citet{Hoppa99, porco2006cassini}). On Pluto, however, tidal forces are presently insignificant because the Pluto/Charon system has reached the end-point of its tidal evolution, a dual-synchronous state. Another mechanism for driving tectonic activity in icy bodies is global volume change due to phase changes in the ice shell \citep{MuellerMcKinnon}. If Pluto has an ocean that is slowly freezing, it would cause global volume expansion and extensional stresses at the surface \citep{robuchon2011thermal}. Conversely, ice Ih converting into a denser phase, such as ice II, would lead to global volume contraction and compressional stresses \citep{mckinnon1997composition}. The dominance of extensional tectonic features across Pluto's surface could be explained by the freezing of a subsurface ocean. But would such an ocean have frozen billions of years ago or might it still exist today?

Previous models of Pluto's thermal evolution by \citet{robuchon2011thermal} found that if Pluto's ice shell transmits heat conductively, an ocean forms and likely persists to the present. However, the persistence of a subsurface ocean may be sensitive to the thickness of Pluto's H$_2$O layer, the thermal conductivity of the silicate core, and the thermal conductivity of the ice shell. \citet{robuchon2011thermal} used a constant thermal conductivity for ice of $k_i=2.26$ W$/$m$/$K and did not treat the formation of ice II. However, the thermal conductivity of ice is strongly temperature dependent \citep{PW}. Higher thermal conductivities would allow the ice shell to cool more efficiently, leading to colder final states and possibly, the formation of ice II. Additionally, silicate conductivity is one of the most important parameters influencing the thermal evolution of ice-rock bodies, yet many planetary models do not provide strong justification for their chosen values \citep{desch2009thermal}. The silicate conductivity in such models vary from $1$ W$/$m$/$K \citep{desch2009thermal}, to 4.2 W$/$m$/$K \citep{robuchon2011thermal}. 

We use numerical models to re-examine the thermal evolution of Pluto to test the hypothesis that the partial freezing of an ocean could be driving tectonic activity. We vary the physical properties of both the silicate core and the ice shell to identify the conditions under which a subsurface ocean will survive, and conditions that lead to the formation of ice II. 
\section{Methods}
\subsection{Initial Conditions}

We simulate the thermal evolution of Pluto using a one-dimensional, explicit finite-difference model,
\begin{equation}
\rho C_p \frac {dT}{dt}= \frac{H}{\rho}+\frac{1}{r^2}\frac{1}{dr}\big{(}r^2 k \frac{dT}{dr}\big{)},
\end{equation}
where $T$ is temperature, $t$ is time, $r$ is radial position, $C_p$ is the specific heat, $\rho$ is density, $H$ is the heating per unit mass, and $k$ is thermal conductivity. The value of each of these variables, explained below, depend on temperature, time, and depth. We use a node spacing of 2 km and a time step of less than 1000 years, keeping the Courant number below 0.25. 

We use a constant surface temperature of $T_s = 40$ K, appropriate for the long-term average surface temperature of Pluto \citep{Tryka1994temperature}, and an initially homogeneous internal temperature $T_0$, which we vary between $150 - 250$ K. The initial temperature does not affect the final thermal state because radiogenic heating dominates the thermal evolution \citep{robuchon2011thermal}. The Pluto-Charon system likely formed in a giant impact \citep{canup2005giant, Canup2011}. We assume Pluto becomes tidally locked early in its history ($10 - 100$ Myr after the impact) \citep{dobrovolskis1997dynamics, barr2015tectonic}. We begin our simulations just after tidal evolution is complete, at $t_0 = 100$ Myr after CAI formation.
\subsection{Silicate Core}

We assume Pluto is fully differentiated into a silicate core and a water ice layer; the energy released by accretion and the decay of radioactive isotopes early in Pluto's history should be more than sufficient to melt water ice and effectively separate icy and rocky components \citep{mckinnon1997composition, Hussmann2006, mckinnon2008structure, desch2009thermal, robuchon2011thermal}. The density of silicates appropriate for Kuiper Belt objects and other ice-rock bodies is uncertain (see e.g., \citealt{mckinnon2008structure} for discussion).  A density of $\rho_r \approx 2.4$ g cm$^{-3}$ has been inferred for the rocky core of Saturn's satellite Enceladus \citep{iess2014gravity}, consistent with a core composed of hydrated silicates such as serpentinite. If silicates avoided hydrothermal alteration, or if they were subsequently dehydrated, Pluto's core may be composed primarily of olivine and pyroxene \citep{mckinnon2008structure}. A reasonable upper bound for core density is $\rho_r\approx3.5$ g cm$^{-3}$, appropriate for the intrinsic density of carbonaceous chondrites \citep{yomogida1983physical, macke2011density}. The radius of the core, 
\begin{equation}
R_c=R\bigg{(}\frac{\bar{\rho}-\rho_i}{\rho_r-\rho_i}\bigg{)}^{1/3},
\end{equation}
where $\rho_i= 0.92$ g cm$^{-3}$ is the density of ice Ih. This yields an initial ice shell thickness $D = 180$ to $340$ km. 

The core is heated by long lived radioisotopes $^{238}$U, $^{235}$U, $^{232}$Th and $^{40}$K. We assume an initial abundance of parent isotopes appropriate for CI chondrites \citep{lodders2003solar}. The heating per unit mass of rock is then $H(t) = H_0 \chi \exp(-t/\tau)$, where the values for initial heating $H_0$, concentration $\chi$, and half-life $\tau$ are taken from Table 3 of \citet{robuchon2011thermal}. Heat transfer in the core occurs by conduction. The core has a specific heat, $C_{p,r}=800$ J kg$^{-1}$ K$^{-1}$. The thermal conductivity of the core depends on its composition, which is uncertain. Ordinary and carbonaceous chrondrites have silicate conductivities $k_r=1-2$ W m$^{-1}$ K$^{-1}$ \citep{yomogida1983physical}.  The conductivity of antigorite at low pressure is $k_r=2.4$ W m$^{-1}$ K$^{-1}$ \citep{osako2010thermal}, the average conductivity of the terrestrial mantle is $k_r=3$ W m$^{-1}$ K$^{-1}$ \citep{MantleConductivity}, and olivine has a thermal conductivity of $k_r=4$ W m$^{-1}$ K$^{-1}$ \citep{osako2004simultaneous}. Hydrothermal circulation, if present, could increase the effective thermal conductivity of the core by advecting heat toward the surface (\textit{e.g.} \citet{neveu2015core}). Because the composition and thermal properties of the core are uncertain, we vary the conductivity from $k_r=1$ to $4$ W m$^{-1}$ K$^{-1}$ to explore how different values affect the thermal evolution. 
\subsection{Ice Shell}
 
It is important to consider whether Pluto's ice shell is likely to undergo solid state convection. We calculate the Rayleigh Number $Ra$, a dimensionless parameter that describes the vigor of convection, and compare it to the critical value for convection in the stagnant lid regime \citep{Solomatov95}. The critical value for convection is defined as \citep{Solomatov95} 
\begin{equation}
Ra_{cr}=20.9\bigg{(}\frac{Q^*(T_b-T_s)}{RT_i^2}\bigg{)}^4
\end{equation}
where $Q^*=59.4$ kJ/mol is the activation energy for diffusion in ice \citep{GoldsbyKohlstedt}, $R$ is the gas constant, $T_b$ is the temperature at the base of the ice shell and $T_i$ is the average temperature of the convecting layer which we approximate as $T_i=T_b$ \citep{mckinnon2006convection}. We calculate the critical Rayleigh number for convection in the ice shell to be $Ra_{cr}=7\times10^6$. 

The Rayleigh number is defined as
\begin{equation}
Ra = \frac{\rho_ig\alpha(T_b-T_s)D^3}{\kappa\eta_i},
\end{equation}
where $\alpha$ is the coefficient of thermal expansion,  $\kappa=k_i/(\rho_iC_{p,i})$ is the thermal diffusivity of ice and $\eta_i$ is the viscosity of the ice evaluated at $T_i$. Under the low stress conditions expected in the ice shell, any plastic deformation will occur via diffusion creep and the viscosity $\eta_i=Ad^2exp(Q/RT_i)$, where $d$ is grain size and $A=8.36\times10^9 (T_i/250$ K$)$ Pa s/m$^2$ is an experimentally constrained constant \citep{GoldsbyKohlstedt, barr2008}. The ice grain size in convecting ice shells likely ranges from $ 1-100$ mm, depending on whether grain growth is limited by pinning from impurities or by dynamic recrystalization \citep{barrMcKinnon2007}. We calculate the Rayleigh number evaluating the physical properties of ice at $T_i=250$ K, and the effects of spherical geometry and variable conductivity on the Rayleigh number are discussed in the supplemental material (S2).

We find that the ice shell is unlikely to convect and that the Rayleigh number only exceeds the critical value for the thickest possible ice shells, $D=340$ km, with the smallest possible grain sizes, $d<2 $ mm, when the viscosity of the ice $\eta_i<6\times10^{16}$ Pa s.  Our findings are consistent with \citet{robuchon2011thermal} who find that the ice shell will convect only if $\eta<= 10^{16}$ Pa s. We therefore assume heat transport in the ice shell occurs by thermal conduction and do not model convection. We use a temperature dependent specific heat $C_{p,i} = 8.8 T$ J kg$^{-1}$ K$^{-2}$ \citep{PW} and temperature dependent thermal conductivity of $k_i=488/T +0.48$ W$/$m$/$K \citep{Hobbs1974}. The conductivity near the surface is likely reduced due to porosity and the presence of volatile ices such as nitrogen, therefore, in the top 10 km we use a constant thermal conductivity of $ k_i=1$ W m$^{-1}$ K$^{-1}$ \citep{shoshany2002monte}.

\subsection{Phase Changes in H$_2$O Layer}
If the heat flow out of the core is larger than the maximum heat flow capable of moving through the ice shell, melting at the base of the ice shell can occur. The heat flow into the ocean boundary, if the boundary is at element $i$, is $F_{in}=F_{core}(R_c/r)^2$, where $F_{core}$ is the heat flow from the core and $r$ is the radial position of the ocean boundary. The heat flow away from the boundary is $F_{out}=\big(\frac{k_{i}+k_{i-1}}{2}\big)\frac{T_{i}-T_{i-1}}{\Delta r}$. Over a time step $\Delta t$, the boundary moves a distance $\Delta z= \Delta t \frac{F_{in}-F_{out}}{\rho L}$, where $L$ is the latent heat. When $ \Delta z > \Delta r$, we consider the phase transition in that element complete. We treat the transformation between ice Ih to liquid and ice Ih to ice II in this way. We benchmarked our code against an exact solution to the classical Stefan problem (see Supplemental Figure S1). We have also benchmarked our code against the results of \citep{robuchon2011thermal} for the case of a conductive ice shell (S2). 

We assume the ocean maintains a constant temperature, $T_m$, with depth, because the adiabatic temperature gradient is small \citep{melosh2004temperature}. The melting temperature of ice depends on pressure, in addition to the presence of volatiles such as salts and ammonia. A small amount of ammonia can significantly reduce the liquidus of the water-ammonia system \citep{kargel1992ammonia}. Some Kuiper Belt accretion models suggest that Pluto's ice shell might be 5 percent ammonia by mass \citep{mckinnon2008structure}. This would reduce the liquidus to $T_m=250$ K. We use a melting temperature that depends on pressure and ammonia concentration following \citet{LK} and assume a constant mass fraction of ammonia in the H$_2$0 layer, $X$, between $0$, $5$ and $10\%$. As the ice-ocean boundary changes depth, the melting temperature changes due to changes in lithostatic pressure. For example, if the ocean is growing, the melting temperature increases as pressures at the ocean boundary decreases. The ocean must therefore warm up in order to reach this new melting temperature. This energy is accounted for in our model. 

We also treat the phase transition between ice Ih and ice II. While ice Ih has a hexagonal crystal structure, ice II is a dense polymorph of ice with a rhombohedral structure and a density $\rho=1.18$ g cm$^{-3}$ \citep{Hobbs1974}. Pre-encounter interior models by \citet{mckinnon1997composition} suggest that ice II is likely only present in Pluto's interior if Pluto is undifferentiated, because for a differentiated body, the pressure at the base of the ice shell may not be high enough for ice II to form. However, Pluto's more ice-rich composition revealed by \emph{New Horizons} data suggests that ice II could form in a differentiated Pluto. We account for the temperature dependence of the ice Ih to ice II transition. The equilibrium temperature, $T_{eq}$, between these phases is,
\begin{equation} 
T_{eq}=\frac{P-14}{0.827},
\end{equation}
where $P$ is pressure in MPa, and temperature is in units of Kelvin \citep{bridgman1912water,whalley1984energies, leon2002thermodynamics}. As temperature drops, ice II becomes the stable phase in the ice shell. For likely pressures at the base of Pluto's shell, ice II begins to form when temperatures drop below $\sim230$ K. The conversion of ice Ih to ice II liberates $\sim$1 kJ mol$^{-1}$ \citep{handa1988energies}. Pressure is approximated as $P=\rho_{i}g_{z}z$, where $z$ is depth, $g_{z}=(GM_r)/r^2$ is the depth-dependent gravity, $G$ is the gravitational constant and $M_r$ is the mass below the local radius $r$. Gravity increases slightly from $g=0.67$ m s$^{-2}$ at the surface to $g\approx0.75$ m s$^{-2}$ at the base of the ice shell. \citet{mckinnon2005ice} noted that ice XI may be present on the surface of Pluto, but we do not treat this phase because its formation is not associated with any significant volume change \citep{Hobbs1974}. 
\subsection{Global Volume Change} 
Phase changes in the ice shell and thermal expansion can both contribute to global volume change. At each time step, we measure the volume change in each element due to these effects and calculate the change in radius, $\Delta R$, and the surface strain, $\epsilon = \Delta R/R$. We use a constant coefficient of thermal expansion of $\alpha=10^{-4}$ K$^{-1}$ in the ice shell and $\alpha=10^{-5}$ K$^{-1}$ in the core. The strain from thermal expansion is over an order of magnitude less than from H$_2$O phase changes. The stress resulting from global volume change is \citep{melosh1977global},
\begin{equation}
\sigma=2\mu\big{(}\frac{1+\nu}{1-\nu}\big{)}\epsilon,
\end{equation}
where $\mu=3.5 $ GPa is the shear modulus of ice and $\nu=0.33$ is Poisson's ratio \citep{gammon1983elastic}. To account for brittle failure at the surface, we use a plastic failure strength of $-10$ MPa in compression and $1$ MPa in tension \citep{schulson1999structure}, and the stress in the ice shell remains confined between these two values. 

\section{Results}
\begin{figure}
\centering{\includegraphics[width=26pc]{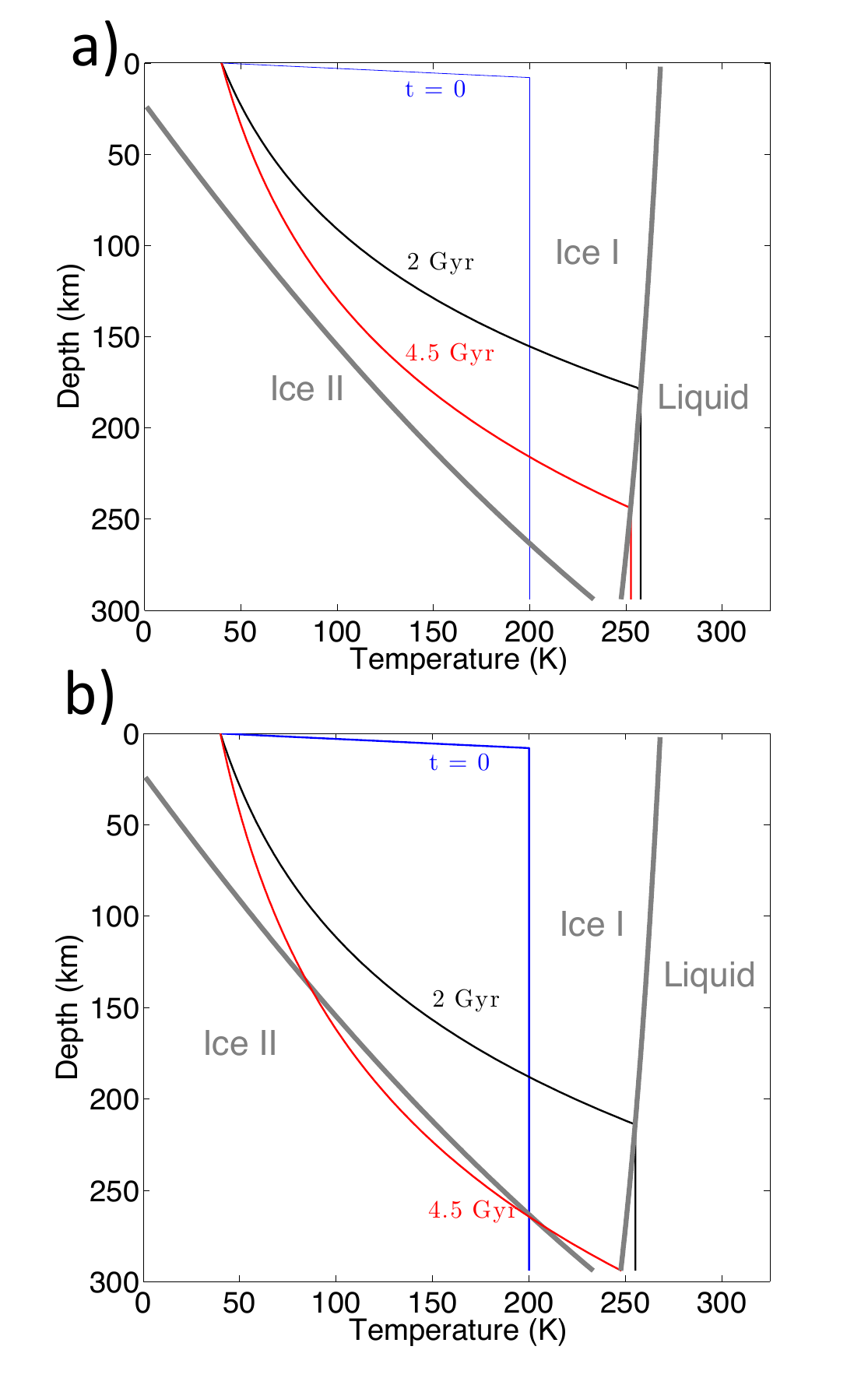}}
 \caption{ Temperature profile in the ice shell of Pluto. Blue lines show the initial temperature condition. Black lines and red lines show the temperature at $2$ Gyr and $4.5$ Gyr after CAI formation. Thick gray lines show the phase boundary between ice Ih and ice II, and between ice Ih and liquid. The simulations shown have a mass fraction of $5\%$ ammonia in the ice shell. Panel (a) shows a simulation with core density $\rho_r=3.05$ g cm$^{-3}$, ice shell thickness $D=285$ km, and silicate thermal conductivity $k_r=2.5$ W m$^{-1}$ K$^{-1}$. In this case, the ocean persists to present day and no ice II forms. Panel (b) shows a simulation with $\rho_r=3.05$ g cm$^{-3}$, $D=285$ km and $k_r=3.5$ W m$^{-1}$ K$^{-1}$. In this simulation, the ocean freezes and the base of the ice shell cools such that 100 km of ice II forms.   \label{fig:figure1}}
 \end{figure}

We simulate Pluto's thermal evolution for a wide range of silicate core properties and ice shell properties. We find that Pluto either reaches a ``warm'' final state or a ``cold'' final state depending on the thermal properties of the silicate core. Figure \ref{fig:figure1}a shows an example of a warm final state, where an ocean forms and survives to the present day. Here, $k_{r}=2.5$ W m$^{-1}$ K$^{-1}$ and $\rho_{r}=3.05$ g cm$^{-3}$. The ocean would currently be $50$ km thick, beneath $250$ km of ice. The simulations shown here have a nominal ammonia concentration of $5\%$. We find that varying the ammonia concentration in the ice shell changes the ocean thickness by $10-20$ km (see S3). 

Figure \ref{fig:figure1}b shows the results of a simulation with $k_{r}=3.5$ W m$^{-1}$ K$^{-1}$ and $\rho_{r}=3.05$ g cm$^{-3}$, where Pluto reaches a cold final state. An ocean forms early in Pluto's history but freezes completely by $t=4.5$ Gyr. As the last few kilometers of the ocean start to freeze, ice II begins to form. The curved temperature profile through the ice shell, which results from the temperature dependent thermal conductivity of ice, causes more of the ice shell to be at colder temperatures and therefore increases the likelihood that ice II will form. 

\begin{figure}
  \centering{\includegraphics[width=40pc]{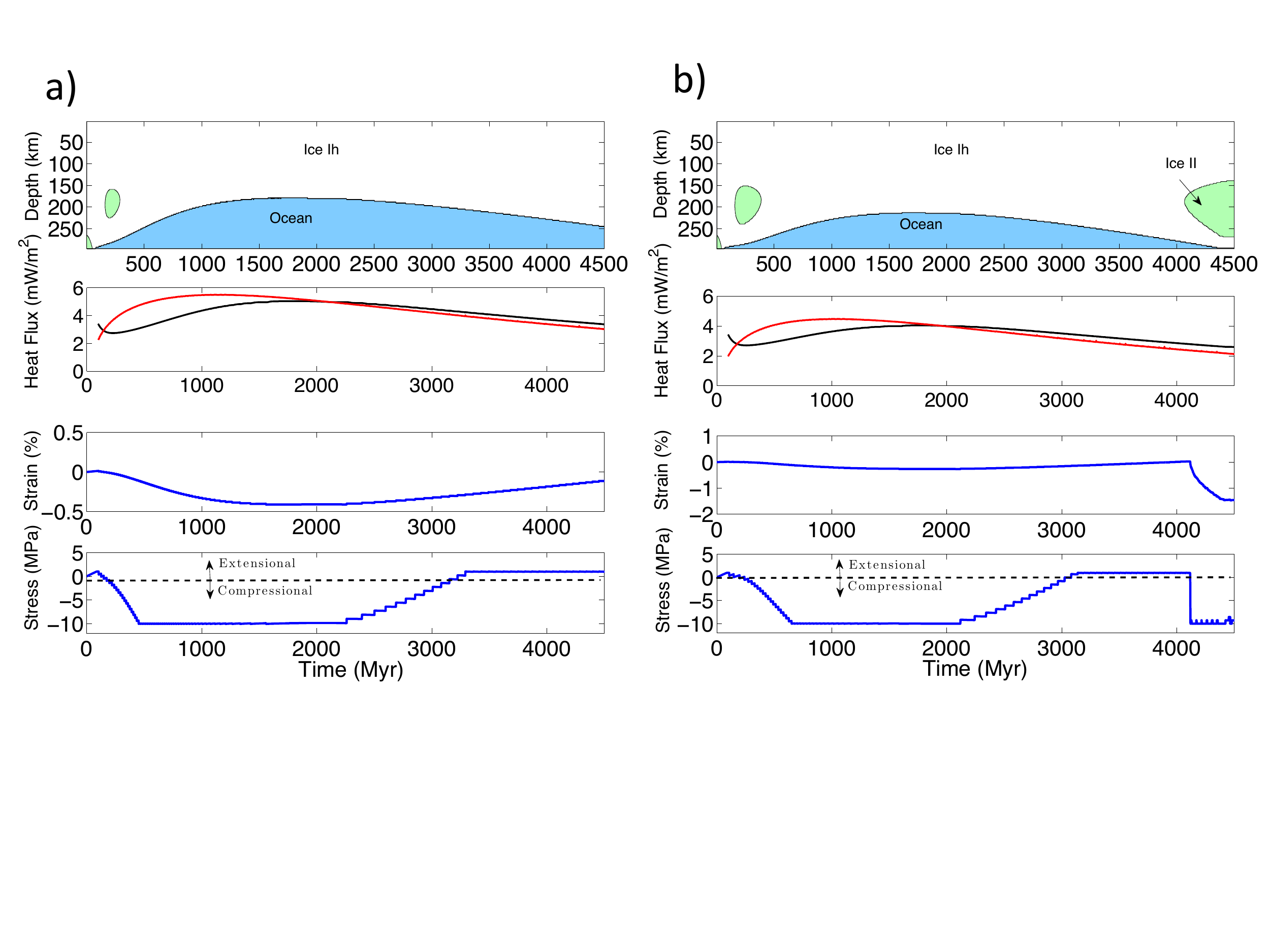}}
 \caption{The evolution of the ice shell through time, with corresponding stress and strain at the surface, for the two cases shown in Figure \ref{fig:figure1}, (left panels correspond to Figure \ref{fig:figure1}a, right panels correspond to Figure \ref{fig:figure1}b). The top panel shows a cross section of the ice shell through time, with depth on the $y$-axis and time after CAI formation on the $x$-axis. Color corresponds to H$_2$O phase,with white, green and blue representing ice Ih, ice II, and liquid, respectively. The second panel from the top shows the heat flow out of the core (red line) and the surface heat flux (black line). The bottom center panel shows the strain at the surface as the result of phase changes in the ice shell and thermal expansion. The bottom panel shows stress at the surface. Compressive stresses are negative and extensional are positive.
 \label{fig:figure2}}
 \end{figure}

Figure \ref{fig:figure2} illustrates the phase, strain, and stress evolution of Pluto's H$_2$O layer for both the ``warm history'' and ``cold history'' depicted in Figure \ref{fig:figure1}. In both cases, a layer of ice II forms early on due to our assumed initial temperature in Pluto's interior, $T_0=200$ K, but the ice II soon disappears after Pluto warms up. The growth of the ocean leads to compressional stresses, which reverse as the ocean begins to freeze after $t\sim 2$ Gyr. In the warm scenario, (Figure \ref{fig:figure2}b), the slow freezing of the ocean from a thickness of $100$ to $50$ km causes an extensional surface strain of $0.3\%$ and generates extensional surface stresses exceeding the tensile failure stress of ice. Contraction due to the cooling of the ice shell is small compared to the expansion from ocean freezing. Extensional failure occurs into the present day. In the cold scenario, however, the ocean freezes and ice II begins to form in the last $300$ Myr. The formation of ice II results in a surface contraction of about $1.5\%$ (Figure \ref{fig:figure2}b), during which surface stresses reach the compressive failure stress. 

Our results illustrate that the evolution of the H$_2$O layer is sensitive to the thermal properties of the core. Figure 3 shows how the present state of the ice shell depends on the density and conductivity of the silicate core. Silicate conductivities less than $k_r \sim 3$ W m$^{-1}$ K$^{-1}$ allow Pluto to remain warm enough to sustain a subsurface ocean and preclude the formation of ice II. Higher silicate conductivities allow the core to lose heat more efficiently, leading to scenarios where the ocean freezes and ice II forms. However for ice shells thinner than $260$ km, (corresponding to silicate densities less than $2.9$ g$/$cm$^3$), the pressures at the base of the ice shell are not high enough to form ice II, so the ocean may freeze without forming ice II. In the case of high core density and very high conductivity, the ice II layer can reach a maximum thickness of 200 km. 

\section{Discussion}
Our results show that if the ocean freezes, and if the ice shell is thicker than $260$ km, then ice II will form. The recent formation of ice II would lead to global volume contraction and compressional tectonic activity at the surface. Since there is no strong evidence of compressional tectonic activity, \citep{moore2016geology}, we conclude that ice II has not formed. 

Ice II formation can be prevented either by keeping Pluto warm enough to sustain a subsurface ocean, or if the silicate core density is less than $2.9$ g$/$cm$^3$ and the ice shell is less than $260$ km thick. There are several reasons that a thicker ice shell with a subsurface ocean may be more likely than a thin ice shell. We find that the temperatures in the inner $750$ km of the silicate core exceed $850$ K, the temperature where low density hydrous silicates destabilize \citep{ulmer1995serpentine}. Therefore at least $60\%$ of the core by volume is likely dehydrated into olivine and pyroxene with $\rho_r=3.5$ g$/$cm$^3$, giving a minimum average core density of $3.05$ g$/$cm$^3$. In this case, the minimum H$_2$0 layer thickness is $\sim300$ km. Additionally, the influence of volatile ices such as nitrogen and methane may be more effective at insulating the ocean than shown in our model. We assume volatiles are concentrated in the top 10 km of Pluto's ice shell, but if methane clathrates are abundant in the entire ice shell, its thermal conductivity may be significantly reduced, \citep{mckinnon1997composition, waite2002thermal}, increasing the likelihood that the ocean will survive. We find that if the ice shell has a constant thermal conductivity of $3$ W$/$m$/$K, a subsurface ocean survives even if the thermal conductivity of the core is high (S4). The likelihood of ocean survival further increases when considering that as the ocean begins to freeze, impurities are excluded from the ice shell and ammonia and salt concentrations in the ocean will increase, further reducing the melting temperature.

If Pluto still has an ocean, then global volume expansion would still be occurring, suggesting extensional tectonic activity could be ongoing. This is consistent with the observations of \citet{moore2016geology} who suggest that the relative youth of tectonic features could be consistent with the partial freezing of an ocean. Additionally, tensional stresses at the surface and ocean pressurization, due to the ice shell thickening, could make cryovolcanic resurfacing more likely \citep{robuchon2011thermal}.

The results of our model differ from that of \citet{robuchon2011thermal} in several important ways. First, we treat the formation of ice II, which is important because we find that if the ocean freezes then ice II is likely to form and generate large compressional strains at the surface. Secondly, we use a temperature dependent thermal conductivity in the ice shell, which causes the thermal gradient to increase with depth. We find that this increases the rate at which the ocean freezes, because heat can be transported more efficiently away from the ice-ocean boundary. Lastly, we limit the stress at Pluto's surface to account for brittle failure. This affects the timing of when the stress state switches between compression and tension. We find that in cases where the ocean survives, ocean freezing starts generating extensional failure by $1$Gyr ago and that extensional tectonics should dominate, whereas \citet{robuchon2011thermal} find that cases with an ocean generate mostly compressional features, with minor recent extensional features. 

The partial freezing of an ocean would produce globally distributed extensional faults with no preferred orientation, although heterogenities in the ice shell or sets of pre-existing fractures could strongly influence fault orientation. Other potential mechanisms for generating tectonic features on Pluto include tidal de-spinning and polar wander \citep{keane2016pluto,nimmo2016loading}. Tidal stresses from despinning will no longer occur after Pluto and Charon reach a dual synchronous state, which likely occurred less than $100$ Myr after the Charon forming impact \citep{barr2015tectonic}. Stresses from polar wander would likely create a distribution of compressional, strike-slip and  extensional tectonic features with orientations that could be predicted based on the reorientation angle \citep{leith1996there}. Global surface expansion may be the best explanation for the dominance of extensional tectonic features observed on Pluto, although all of these processes could be important in determining Pluto's complete tectonic evolution. 

\section{Conclusion}

We have shown that while the survival of the ocean depends strongly on the thermal properties of the silicate core, the lack of obvious compressional tectonic features on Pluto's surface suggests that ice II has not formed. Therefore Pluto either has an ocean today or the ice shell is thinner than $260$ km. New Horizons data can be used to distinguish between these hypotheses by characterizing the global geologic stratigraphy, assessing the timing of tectonic deformation events, and searching for evidence of present day extensional tectonic activity.

\section{Acknowledgements} We thank Francis Nimmo and an anonymous reviewer for their helpful comments which have greatly improved the paper. This work is supported by NESSF NNX13AN99H. Author Barr acknowledges support from NASA PG\&G NNX15AN79G. Supporting data and information on our model are available in an SI file.

\clearpage
 
 \clearpage
 
 \clearpage
 \begin{figure}
 \centering{\includegraphics[width=35pc]{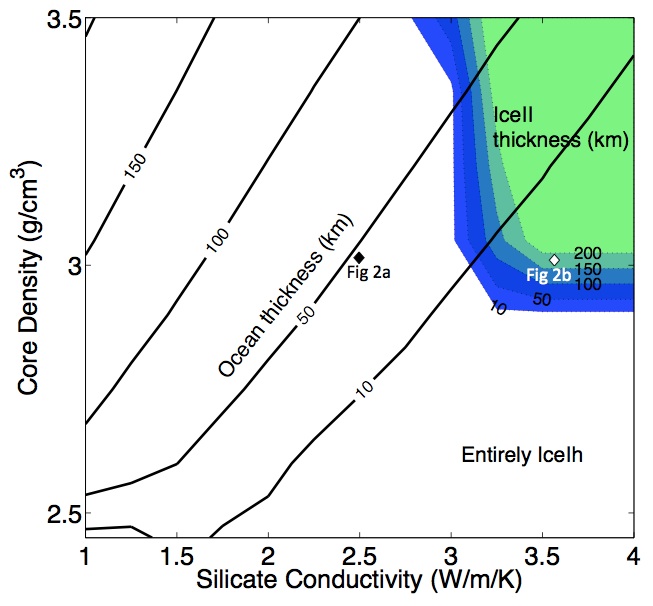}}
 \caption{Present state of Pluto's H$_2$O layer, which depends on the silicate core conductivity ($x$-axis) and density ($y$-axis).  We simulate Pluto's thermal evolution for more than 100 pairs of conductivity and density values.  Black lines show contours for final ocean thickness. Blue and green shaded regions show the final ice II thickness. This suite of simulations used an ammonia concentration of $X=5$ \% by mass. White and black diamonds indicate where the final state of the ice shell plots for the simulations discussed in figures 1 and 2. 
 \label{RadiusChange.pdf}}
 \end{figure}


\newpage
\begin{thebibliography}{48}
\providecommand{\natexlab}[1]{#1}
\expandafter\ifx\csname urlstyle\endcsname\relax
  \providecommand{\doi}[1]{doi:\discretionary{}{}{}#1}\else
  \providecommand{\doi}{doi:\discretionary{}{}{}\begingroup
  \urlstyle{rm}\Url}\fi

\bibitem[{\textit{Barr}(2008)}]{barr2008}
Barr, A.~C. (2008), Mobile lid convection beneath {E}nceladus' south polar
  terrain, \textit{Journal of Geophysical Research: Planets}, \textit{113}(E7).

\bibitem[{\textit{Barr and Collins}(2015)}]{barr2015tectonic}
Barr, A.~C., and G.~C. Collins (2015), Tectonic activity on {P}luto after the
  {C}haron-forming impact, \textit{Icarus}, \textit{246}, 146--155.

\bibitem[{\textit{{Barr} and {McKinnon}}(2007)}]{barrMcKinnon2007}
{Barr}, A.~C., and W.~B. {McKinnon} (2007), Convection in ice {I} shells and
  mantles with self-consistent grain size, \textit{J. Geophys. Res.},
  \textit{112}, E02012, \doi{10.1029/2006JE002781}.

\bibitem[{\textit{Bridgman}(1912)}]{bridgman1912water}
Bridgman, P.~W. (1912), Water, in the liquid and five solid forms, under
  pressure, in \textit{Proceedings of the American Academy of Arts and
  Sciences}, pp. 441--558, JSTOR.

\bibitem[{\textit{Canup}(2005)}]{canup2005giant}
Canup, R.~M. (2005), A giant impact origin of {P}luto-{C}haron,
  \textit{Science}, \textit{307}(5709), 546--550.

\bibitem[{\textit{Canup}(2011)}]{Canup2011}
Canup, R.~M. (2011), On a giant impact origin of {C}haron, {N}ix, and {H}ydra,
  \textit{The Astronomical Journal}, \textit{141}, 35,
  \doi{10.1088/0004-6256/141/2/35}.

\bibitem[{\textit{Desch et~al.}(2009)\textit{Desch, Cook, Doggett, and
  Porter}}]{desch2009thermal}
Desch, S.~J., J.~C. Cook, T.~Doggett, and S.~B. Porter (2009), Thermal
  evolution of {K}uiper {B}elt objects, with implications for cryovolcanism,
  \textit{Icarus}, \textit{202}(2), 694--714.

\bibitem[{\textit{Dobrovolskis et~al.}(1997)\textit{Dobrovolskis, Peale, and
  Harris}}]{dobrovolskis1997dynamics}
Dobrovolskis, A.~R., S.~J. Peale, and A.~W. Harris (1997), Dynamics of the
  {P}luto--{C}haron binary, \textit{{P}luto and {C}haron}, \textit{159}.

\bibitem[{\textit{Gammon et~al.}(1983)\textit{Gammon, Kiefte, and
  Clouter}}]{gammon1983elastic}
Gammon, P., H.~Kiefte, and M.~Clouter (1983), Elastic constants of ice samples
  by brillouin spectroscopy, \textit{The Journal of Physical Chemistry},
  \textit{87}(21), 4025--4029.

\bibitem[{\textit{{Goldsby} and {Kohlstedt}}(2001)}]{GoldsbyKohlstedt}
{Goldsby}, D.~L., and D.~L. {Kohlstedt} (2001), {Superplastic deformation of
  ice: Experimental observations}, \textit{J. Geophys. Res.}, \textit{106},
  11,017--11,030.

\bibitem[{\textit{Handa et~al.}(1988)\textit{Handa, Klug, and
  Whalley}}]{handa1988energies}
Handa, Y.~P., D.~Klug, and E.~Whalley (1988), Energies of the phases of ice at
  low temperature and pressure relative to ice {I}h, \textit{Canadian journal
  of chemistry}, \textit{66}(4), 919--924.

\bibitem[{\textit{{Hobbs}}(1974)}]{Hobbs1974}
{Hobbs}, P.~V. (1974), \textit{{Ice physics}}, Oxford: Clarendon Press.

\bibitem[{\textit{Hofmeister}(1999)}]{MantleConductivity}
Hofmeister, A. (1999), Mantle values of thermal conductivity and the geotherm
  from phonon lifetimes, \textit{Science}, \textit{283}(5408), 1699--1706.

\bibitem[{\textit{{Hoppa} et~al.}(1999)\textit{{Hoppa}, {Tufts}, {Greenberg},
  and {Geissler}}}]{Hoppa99}
{Hoppa}, G.~V., B.~R. {Tufts}, R.~{Greenberg}, and P.~E. {Geissler} (1999),
  {Formation of cycloidal features on Europa}, \textit{Icarus}, \textit{141},
  287--298.

\bibitem[{\textit{{Hussmann} et~al.}(2006)\textit{{Hussmann}, {Sohl}, and
  {Spohn}}}]{Hussmann2006}
{Hussmann}, H., F.~{Sohl}, and T.~{Spohn} (2006), {Subsurface oceans and deep
  interiors of medium-sized outer planet satellites and large trans-neptunian
  objects}, \textit{Icarus}, \textit{185}, 258--273,
  \doi{10.1016/j.icarus.2006.06.005}.

\bibitem[{\textit{Iess et~al.}(2014)}]{iess2014gravity}
Iess, L., et~al. (2014), The gravity field and interior structure of
  {E}nceladus, \textit{Science}, \textit{344}(6179), 78--80.

\bibitem[{\textit{Kargel}(1992)}]{kargel1992ammonia}
Kargel, J.~S. (1992), Ammonia-water volcanism on icy satellites: Phase
  relations at 1 atmosphere, \textit{Icarus}, \textit{100}(2), 556--574.

\bibitem[{\textit{Keane and Matsuyama}(2016)}]{keane2016pluto}
Keane, J., and I.~Matsuyama (2016), {P}luto followed its heart: True polar
  wander of {P}luto due to the formation and evolution of {S}putnik {P}lanum,
  in \textit{Lunar and Planetary Science Conference}, vol.~47, p. 2348.

\bibitem[{\textit{Leith and McKinnon}(1996)}]{leith1996there}
Leith, A.~C., and W.~B. McKinnon (1996), Is there evidence for polar wander on
  europa?, \textit{Icarus}, \textit{120}(2), 387--398.

\bibitem[{\textit{{Leliwa-Kopysty{\' n}ski}
  et~al.}(2002)\textit{{Leliwa-Kopysty{\' n}ski}, {Maruyama}, and
  {Nakajima}}}]{LK}
{Leliwa-Kopysty{\' n}ski}, J., M.~{Maruyama}, and T.~{Nakajima} (2002), {The
  water-ammonia phase diagram up to 300 MPa: Application to icy satellites},
  \textit{Icarus}, \textit{159}, 518--528.

\bibitem[{\textit{Leon et~al.}(2002)\textit{Leon, Romo, and
  Tchijov}}]{leon2002thermodynamics}
Leon, G.~C., S.~R. Romo, and V.~Tchijov (2002), Thermodynamics of high-pressure
  ice polymorphs: ice ii, \textit{Journal of Physics and Chemistry of Solids},
  \textit{63}(5), 843--851.

\bibitem[{\textit{Lodders}(2003)}]{lodders2003solar}
Lodders, K. (2003), Solar system abundances and condensation temperatures of
  the elements, \textit{The Astrophysical Journal}, \textit{591}(2), 1220.

\bibitem[{\textit{Macke et~al.}(2011)\textit{Macke, Consolmagno, and
  Britt}}]{macke2011density}
Macke, R.~J., G.~J. Consolmagno, and D.~T. Britt (2011), Density, porosity, and
  magnetic susceptibility of carbonaceous chondrites, \textit{Meteoritics \&
  Planetary Science}, \textit{46}(12), 1842--1862.

\bibitem[{\textit{McKinnon and Hofmeister}(2005)}]{mckinnon2005ice}
McKinnon, W., and A.~Hofmeister (2005), Ice xi on {P}luto and {C}haron?, in
  \textit{Bulletin of the American Astronomical Society}, vol.~37, p. 732.

\bibitem[{\textit{McKinnon et~al.}(1997)\textit{McKinnon, Simonelli, and
  Schubert}}]{mckinnon1997composition}
McKinnon, W., D.~Simonelli, and G.~Schubert (1997), Composition, internal
  structure, and thermal evolution of {P}luto and {C}haron, \textit{{P}luto and
  {C}haron}, \textit{1}, 295.

\bibitem[{\textit{McKinnon}(2006)}]{mckinnon2006convection}
McKinnon, W.~B. (2006), On convection in ice i shells of outer solar system
  bodies, with detailed application to callisto, \textit{Icarus},
  \textit{183}(2), 435--450.

\bibitem[{\textit{McKinnon et~al.}(2008)\textit{McKinnon, Prialnik, Stern, and
  Coradini}}]{mckinnon2008structure}
McKinnon, W.~B., D.~Prialnik, S.~A. Stern, and A.~Coradini (2008), Structure
  and evolution of {K}uiper {B}elt objects and dwarf planets, \textit{The Solar
  System Beyond Neptune}, \textit{1}, 213--241.

\bibitem[{\textit{McKinnon et~al.}(2016)}]{McKinnon2016convection}
McKinnon, W.~B., et~al. (2016), Convection in a volatile nitrogen-ice-rich
  layer drives {P}luto's geological vigour, \textit{Nature}, \textit{534}(7605),
  82--85.

\bibitem[{\textit{Melosh et~al.}(2004)\textit{Melosh, Ekholm, Showman, and
  Lorenz}}]{melosh2004temperature}
Melosh, H., A.~Ekholm, A.~Showman, and R.~Lorenz (2004), The temperature of
  {E}uropa's subsurface water ocean, \textit{Icarus}, \textit{168}(2),
  498--502.

\bibitem[{\textit{Melosh}(1977)}]{melosh1977global}
Melosh, H.~J. (1977), Global tectonics of a despun planet, \textit{Icarus},
  \textit{31}(2), 221--243.

\bibitem[{\textit{Moore et~al.}(2016)}]{moore2016geology}
Moore, J.~M., et~al. (2016), The geology of {P}luto and {C}haron through the
  eyes of {N}ew {H}orizons, \textit{Science}, \textit{351}(6279), 1284--1293.

\bibitem[{\textit{{Mueller} and {McKinnon}}(1988)}]{MuellerMcKinnon}
{Mueller}, S., and W.~B. {McKinnon} (1988), {Three-layered models of Ganymede
  and Callisto - Compositions, structures, and aspects of evolution},
  \textit{Icarus}, \textit{76}, 437--464, \doi{10.1016/0019-1035(88)90014-0}.

\bibitem[{\textit{Neveu et~al.}(2015)\textit{Neveu, Desch, and
  Castillo-Rogez}}]{neveu2015core}
Neveu, M., S.~J. Desch, and J.~C. Castillo-Rogez (2015), Core cracking and
  hydrothermal circulation can profoundly affect ceres' geophysical evolution,
  \textit{Journal of Geophysical Research: Planets}, \textit{120}(2), 123--154.

\bibitem[{\textit{Nimmo et~al.}(2016)}]{nimmo2016loading}
Nimmo, F., et~al. (2016), Loading, relaxation, and tidal wander at {S}putnik
  {P}lanum, {P}luto, in \textit{Lunar and Planetary Science Conference},
  vol.~47, p. 2207.

\bibitem[{\textit{Osako et~al.}(2004)\textit{Osako, Ito, and
  Yoneda}}]{osako2004simultaneous}
Osako, M., E.~Ito, and A.~Yoneda (2004), Simultaneous measurements of thermal
  conductivity and thermal diffusivity for garnet and olivine under high
  pressure, \textit{Physics of the Earth and Planetary Interiors},
  \textit{143}, 311--320.

\bibitem[{\textit{Osako et~al.}(2010)\textit{Osako, Yoneda, and
  Ito}}]{osako2010thermal}
Osako, M., A.~Yoneda, and E.~Ito (2010), Thermal diffusivity, thermal
  conductivity and heat capacity of serpentine (antigorite) under high
  pressure, \textit{Physics of the Earth and Planetary Interiors},
  \textit{183}(1), 229--233.

\bibitem[{\textit{{Petrenko} and {Whitworth}}(1999)}]{PW}
{Petrenko}, V.~F., and R.~W. {Whitworth} (1999), \textit{{Physics of Ice}},
  Oxford University Press, New York.

\bibitem[{\textit{Porco et~al.}(2006)}]{porco2006cassini}
Porco, C., et~al. (2006), Cassini observes the active south pole of
  {E}nceladus, \textit{science}, \textit{311}(5766), 1393--1401.

\bibitem[{\textit{Robuchon and Nimmo}(2011)}]{robuchon2011thermal}
Robuchon, G., and F.~Nimmo (2011), Thermal evolution of {P}luto and
  implications for surface tectonics and a subsurface ocean, \textit{Icarus},
  \textit{216}(2), 426--439.

\bibitem[{\textit{Schulson}(1999)}]{schulson1999structure}
Schulson, E.~M. (1999), The structure and mechanical behavior of ice,
  \textit{JOM}, \textit{51}(2), 21--27.

\bibitem[{\textit{Shoshany et~al.}(2002)\textit{Shoshany, Prialnik, and
  Podolak}}]{shoshany2002monte}
Shoshany, Y., D.~Prialnik, and M.~Podolak (2002), Monte carlo modeling of the
  thermal conductivity of porous cometary ice, \textit{Icarus},
  \textit{157}(1), 219--227.

\bibitem[{\textit{{Solomatov}}(1995)}]{Solomatov95}
{Solomatov}, V.~S. (1995), {Scaling of temperature- and stress-dependent
  viscosity convection}, \textit{Physics of Fluids}, \textit{7}, 266--274.

\bibitem[{\textit{Stern et~al.}(2015)}]{stern2015pluto}
Stern, S., et~al. (2015), The {P}luto system: Initial results from its
  exploration by new horizons, \textit{Science}, \textit{350}(6258), aad1815.

\bibitem[{\textit{Tryka et~al.}(1994)\textit{Tryka, Brown, Cruikshank, Owen,
  Geballe, and Debergh}}]{Tryka1994temperature}
Tryka, K.~A., R.~H. Brown, D.~P. Cruikshank, T.~C. Owen, T.~R. Geballe, and
  C.~Debergh (1994), Temperature of nitrogen ice on {P}luto and its
  implications for flux measurements, \textit{Icarus}, \textit{112}(2),
  513--527.

\bibitem[{\textit{Ulmer and Trommsdorff}(1995)}]{ulmer1995serpentine}
Ulmer, P., and V.~Trommsdorff (1995), Serpentine stability to mantle depths and
  subduction-related magmatism, \textit{Science}, \textit{268}(5212), 858--861.

\bibitem[{\textit{Waite et~al.}(2002)\textit{Waite, deMartin, Kirby, Pinkston,
  and Ruppel}}]{waite2002thermal}
Waite, W., B.~deMartin, S.~Kirby, J.~Pinkston, and C.~Ruppel (2002), Thermal
  conductivity measurements in porous mixtures of methane hydrate and quartz
  sand, \textit{Geophysical Research Letters}, \textit{29}(24).

\bibitem[{\textit{Whalley}(1984)}]{whalley1984energies}
Whalley, E. (1984), Energies of the phases of ice at zero temperature and
  pressure, \textit{The Journal of chemical physics}, \textit{81}(9),
  4087--4092.

\bibitem[{\textit{Yomogida and Matsui}(1983)}]{yomogida1983physical}
Yomogida, K., and T.~Matsui (1983), Physical properties of ordinary chondrites,
  \textit{Journal of Geophysical Research: Solid Earth}, \textit{88}(B11),
  9513--9533.

\end{thebibliography}
\end{document}